\documentclass[11pt, a4papper]{article}
\usepackage{amsmath,amsfonts,amssymb,amsthm,enumerate,graphicx,xcolor,url}
\usepackage{mathtools}
\usepackage{hyperref}
\usepackage{algorithm}
\usepackage{tikz}
\usepackage{booktabs}
\usepackage{caption}
\usepackage{siunitx}
\usepackage{tabularx}
\usepackage{graphicx}
\usepackage{adjustbox}
\usepackage{multirow}
\usepackage[pagewise]{lineno}
\usetikzlibrary{trees,arrows,positioning,fit,calc}
\tikzset{block/.style={draw, thick, text width=2cm, minimum height=0.75cm, align=center}
}

\newtheorem{theorem}{Theorem}[section]
\newtheorem{lemma}[theorem]{Lemma}
\theoremstyle{definition}
\newtheorem{definition}[theorem]{Definition}
\newtheorem{example}{Example}

\theoremstyle{proposition}

\theoremstyle{remark}
\newtheorem{remark}{Remark}

\theoremstyle{corollary}
\newtheorem{corollary}[theorem]{Corollary}

\numberwithin{equation}{section}
\usepackage{url} 
\usepackage{todonotes}
\usepackage[bottom]{footmisc}
\usepackage{tikz-cd}
\usepackage{tikz}
\usepackage[mathscr]{euscript}
\usepackage{graphicx}
\usepackage{framed}
\graphicspath{{image/}}
\begin{document}

\title{On the exponent of cyclic codes}
\author{ A. K. Bhagat, R. Sarma}

\date{}

\maketitle

\vspace{-12mm}
\begin{center}

\noindent {\small Department of Mathematics,\\ Indian Institute of Technology Delhi,\\ Hauz Khas, New Delhi-110016, India$^{1}$.}

\end{center}

\footnotetext[1]{{\em E-mail addresses:} \url{anujkumarbhagat632@gmail.com} (Anuj Kumar Bhagat), \url{ritumoni407@gmail.com} \\(Ritumoni Sarma).}

\medskip

\hrule

\begin{abstract}
We propose an algorithm to find a lower bound for the number of cyclic codes over any finite field with any given exponent. Besides, we give a formula to find the exponent of BCH codes.
\end{abstract}

\noindent {\small {\em MSC 2010\,:} 94B05, 94B25, 94B60,11T71.\\
\noindent {\em Keywords:} }
cyclic code; BCH code; exponent.
\medskip
\section{Introduction}
For a prime power $q$, denote by $\mathbb{F}_q$ the field that has exactly $q$ elements. Every finite field $\mathbb{F}_q$ has at least one element $\beta$ that satisfies $\mathbb{F}_q\setminus \{0\}=<\beta>$; such elements are called \emph{primitive elements} of $\mathbb{F}_q$. Let gcd$(n,q)=1$. Then $C_i:=\{(i.q^j(\text{mod}\  n)\in\mathbb{Z}_n: j=\mathbb{N}\cup\{0\}\}$ is called the \emph{cyclotomic cosets} of $q$ in $\mathbb{Z}_n$ containing $i$. A \emph{minimal polynomial} of $\beta\in\mathbb{F}_{q^m}$ over $\mathbb{F}_q$ is a non-zero monic polynomial $f(x)\in \mathbb{F}_q[x]$ of least degree such that $f(\beta)=0$. The minimal polynomial of an element of $\mathbb{F}_{q^m}$ over $\mathbb{F}_q$ is irreducible and is unique. The following theorem determines the minimal polynomial of any element in a finite field:
\begin{theorem}\cite{LING}
Let $\beta \in \mathbb{F}_{q^m}$ be a primitive element. If $C_i$ is the cyclotomic coset of $q$ in $\mathbb{Z}_{q^m-1}$ containing $i$. Then $M^{(i)}(x):=\prod_{j\in C_i}(x-\beta^j)$ is the minimal polynomial of $\beta^i$ over $\mathbb{F}_q$.
\end{theorem}
\begin{corollary}\label{samecyclo}\cite{LING}
Let $\beta \in \mathbb{F}_{q^m}$ be a primitive element. Denote by $M^{(s)}(x)$ the minimal polynomial of $\beta^s$ over $\mathbb{F}_q$. Then $M^{(s)}(x)=M^{(t)}(x)$ iff $s$ and $t$ are in the same cyclotomic coset of $q$ in $\mathbb{Z}_{q^m-1}$. 
\end{corollary}
\begin{lemma}\cite{FIELD}\label{Existenceofe}
Let $g(x)\in\mathbb{F}_q[x]$ be a polynomial of positive degree $m$ such that $x\nmid g(x)$. Then for some integer $e$ with $1\leq e\leq q^m-1$, $g(x)$ divides $x^e-1$.
\end{lemma}
\noindent
Lemma \ref{Existenceofe} motivates the following definition.
\begin{definition}\cite{FIELD}
Let $g(x)\in \mathbb{F}_q[x]\setminus \{0\}$. If $x\nmid g(x)$, then the least natural number $e$ such that $g(x)$ divides $x^e-1$ is called the \emph{order} of $g$ and it is denoted by ord$(g)$ or ord$(g(x))$. If $g(x)=x^rf(x)$ for some $r\in\mathbb{N}$ and $f\in\mathbb{F}_q[x]$ with $x\nmid f(x)$, then ord$(g)$ is ord$(f)$.
\end{definition}
\noindent
The following theorems give some properties of the order of irreducible polynomials, their products and their powers:
\begin{theorem}\label{orderirr}\cite{FIELD}
Let $g(x)\in \mathbb{F}_q[x]$ be an irreducible polynomial over $\mathbb{F}_q$ of degree $m$ such that $x\nmid g(x)$. Then ord$(g)$ is equal to the order of any root of $g$ in the multiplicative group $\mathbb{F}_{q^m}^*$.
\end{theorem}
\begin{theorem}\cite{FIELD}\label{degreeirr}
Suppose $d$ is the order of $q$ in $\mathbb{Z}_e^*$. Then the degree of an irreducible polynomial in $\mathbb{F}_q[x]$ of
order $e$ must be equal to $d$.
\end{theorem}
\begin{theorem}\cite{FIELD}\label{orderpower}
Let char$(\mathbb{F}_q)=p$. Suppose $g(x)\in \mathbb{F}_q[x]$ is irreducible such that $x|g(x)$ and ord$(g)=e$. Then ord$(g^k)=ep^t$, where $t$ is the least positive integer with $p^t\geq k$.
\end{theorem}
\begin{theorem}\label{orderproduct}\cite{FIELD}
Let $g_1,g_2,...,g_k\in \mathbb{F}_q[x]$ be pairwise co-prime non-zero polynomials. Then
\begin{equation*}
    \text{ord}(g_1g_2...g_k)=\text{lcm}(\text{ord}(g_1),\text{ord}(g_2),...,\text{ord}(g_k)).
\end{equation*}
\end{theorem}
In \cite{ANNA}, the order of the generator polynomial of a cyclic code $C$ is termed as the exponent of $C$, denoted by exp$(C)$. The authors posed the problem of determining bounds for the number of cyclic codes of given exponent and finding the exponent of BCH and Reed-Solomon codes.\\\\
The remaining sections are as follows. In Section 2, we give bounds for the number of cyclic codes of given exponent. In Section 3, we determine a formula to find the exponent of BCH code and RS code. In Section 4, we conclude the article.
\section{Cyclic codes of given exponent}
Suppose $g(x)$ is a polynomial over $\mathbb{F}_q[x]$. If $n$ is the length of a cyclic code $C$ that $g(x)$ generates, then $g(x)\mid x^n-1$. Note that there are infinitely many choices of $n$. The exponent of such a code depends upon the choice of $g(x)$ and it divides $n$. Conversely, each monic divisor $g(x)$ of $x^n-1\in \mathbb{F}_q[x]$ generates a unique $[n,k]_q$ cyclic code with $k=n-deg(g(x))$.\\ 
Thus, if $e\in \mathbb{N}$ does not divide $n$, then no cyclic code of length $n$ can have exponent $e$. Suppose $e,n\in \mathbb{N}$ are such that $e|n$. Then cyclic codes of exponent $e$ and length $n$ is generated by a divisor of $x^e-1$ of order $e$. Note that not every divisor of $x^e-1$ is of order $e$, for instance, the divisor $(x+1)$ of $x^2-1 \in \mathbb{F}_2[x]$ has order $1$. Thus, if $e$ divides $n$, then to find the number of cyclic codes of exponent $e$ and length $n$, it is sufficient to consider the factorization of $x^e-1$ over $\mathbb{F}_q$. From this discussion, the following remark is clear.
\begin{remark}
The number of cyclic codes of exponent $e$ and length $n$ where $e\mid n$, is independent of choice of $n$.
\end{remark}
\noindent
Throughout the section, we assume that char$(\mathbb{F}_q)=p$ and $e\mid n$.\\
Suppose $x^e-1=\prod_{i=1}^{r}p_i(x)^{f_i}$ is the prime factorization in $\mathbb{F}_q[x]$. Now, with this notation, the obvious upper bound for the number of cyclic codes of exponent $e$ is $\prod_{i=1}^{r}(f_i+1)$. 
\begin{lemma}\label{orderlessthane}
Let $x^e-1=\prod_{i=1}^{r}p_i(x)^{f_i}$ where each $p_i(x)\in \mathbb{F}_q[x]$ is irreducible.. If $f_i>1$, then ord$(p_i(x))<e$.
\end{lemma}
\textbf{Proof.} Suppose ord$(p_i(x))=e$.
Then by theorem \ref{orderpower}, ord$(p_i(x)^{f_i})=ep^t$, where $t$ is the smallest natural number satisfying $p^t\geq f_i$.\\
But ord$(p_i(x)^{f_i})\leq e$ as $p_i(x)^{f_i}$ divides $x^e-1$. Thus we have,
\begin{align*}
    ep^t&\leq e\\
    \implies p^t&\leq 1\\
    \implies f_i&\leq 1
\end{align*}
which contradicts the hypothesis.\hfill $\square$
\begin{lemma}\label{qmodm}
Let $x^e-1=\prod_{i=1}^{r}p_i(x)^{f_i}$ where each $p_i(x)\in \mathbb{F}_q[x]$ is irreducible. Suppose $d$ is the order of $q$ in $\mathbb{Z}_e^*$ and $f_i=1$. Then ord$(p_i(x))=e$ $\Longleftrightarrow$ deg$(p_i(x))=d$.
\end{lemma}
\textbf{Proof.} Follows directly from Theorem \ref{degreeirr} .\hfill $\square$\\
\begin{lemma}\label{for what powers exponent is e}
Let $x^e-1=\prod_{i=1}^{r}p_i(x)^{f_i}$ where each $p_i(x)\in \mathbb{F}_q[x]$ is irreducible. Suppose $f_i>1$. Then for some $a>1,$ ord$(p_i(x)^a)=e$  $\Longleftrightarrow$ $p$ is the only prime divisor of $\frac{e}{\text{ord}(p_i(x))}$. Furthermore, if $p$ is the only prime divisor of $\frac{e}{\text{ord}(p_i(x))}$, then ord$(p_i(x)^a)=e$ whenever $p^{t-1}<a\leq p^t$, where $t=\log_p{\frac{e}{\text{ord}(p_i(x))}}$.
\end{lemma}
\textbf{Proof.} By Lemma \ref{orderlessthane}, ord$(p_i(x))<e$. So, ord$(p_i(x))=d$, where $d\neq e$ and $d\mid e$.\\
For any $a>1$, we have, $e\geq$ord$(p_i(x)^a)=dp^t$ where $t$ is the smallest natural number satisfying $p^t\geq a$.\\
Suppose ord$(p_i(x)^a)=e$ for some $a>1$. Then $\frac{e}{d}=p^t$.\\
Conversely, suppose $\frac{e}{d}$ is a $p$-power, that is, $\frac{e}{d}=p^t$ for some $t\in \mathbb{N}$. Choose $a$ such that $p^{t-1}<a\leq p^t$. Then ord$(p_i(x)^a)=dp^t=e$.\\ This also justifies the second part. \hfill $\square$\\\\
We now give an algorithm to compute a lower bound for the number of cyclic codes of exponent $e$.\\\\
\textbf{Algotithm.}\\\\
\textbf{Step 1}: Factorize $x^e-1$ over $\mathbb{F}_q$.\\Suppose
\begin{equation*}
    x^e-1=\prod_{i=1}^{r}p_i(x)^{f_i}.
\end{equation*}
where each $p_i(x)$ is irreducible over $\mathbb{F}_q[x]$.\\\\
\textbf{Step 2}: If gcd$(q,e)=1$, find $m$=ord$(q)$ in  $\mathbb{Z}_e^*$; otherwise go to \textbf{Step 6}. \\\\
\textbf{Step 3}: Set $M=\{p_i(x): 1\leq i\leq r, f_i=1, \text{deg}(p_i(x))=m\}$ and $t=\#M$.\\\\
\textbf{Step 4}: There are at least $\sum_{j=0}^{r-t}\sum_{i=1}^{t}\binom{t}{i} \binom{r-t}{j}$ cyclic codes of exponent $e$.\\\\
\textbf{Step 5}: STOP.\\\\
\textbf{Step 6:} Set $N=\{p_i(x): 1\leq i\leq r, f_i>1\}$.\\\\
\textbf{Step 7:} $\forall p_i(x)\in N$, find $k_i=\frac{e}{\text{ord}(p_i(x))}$. \\\\
\textbf{Step 8:} Set $S=\{p_i(x)\in N: k_i\ \text{is a}\ p\text{-power}\}$\\\\
\textbf{Step 9:} $\forall p_i(x)\in S$, find $t_i=\log_p{k_i}$.\\\\
\textbf{Step 10:} There are at least $\sum_{i=1}^{\#S}\sum_{j\neq i}(f_j+1)\phi(p^{t_i})$ cyclic codes of exponent $e$.\\\\
\textbf{Step 11}: STOP.\\\\
To prove the algorithm, it it sufficient to prove Step 4 and Step 10 of the algorithm.\\
To prove Step 4, we show that $e$ is the order of $\prod_{i=1}^{l}p_i(x)\prod_{j={l+1}}^{k}p_j(x)$, where $p_i(x)\in M$ and $p_j(x)\notin M$. In fact,
\begin{align*}
    \text{ord}\left(\prod_{i=1}^{l}p_i(x)\right)&=\text{lcm}\{\text{ord}(p_i(x)): 1\leq i\leq l\}\hspace{2cm}(\text{by Theorem \ref{orderproduct}})\\
    &=\text{lcm}\{e: 1\leq i\leq l\}\hspace{3.5cm}(\text{by Lemma \ref{qmodm}})\\
    &=e.
\end{align*}
Thus,
\begin{align*}
    \text{ord}\left(\prod_{i=1}^{l}p_i(x)\prod_{j={l+1}}^{k}p_j(x)\right)
    &=\text{lcm}\left(\text{ord}\left(\prod_{i=1}^{l}p_i(x)\right), \text{ord}\left(\prod_{j={l+1}}^{k}p_j\left(x\right)\right)\right)\hspace{.1cm}(\text{by Theorem \ref{orderproduct}}).\\
    &=e,\ \text{as ord}\left(\prod_{i=1}^{l}p_i(x)\right)=e.
\end{align*}
Noting that there are exactly $t$ elements in $M$, the number of products of the form\\ $\prod_{i=1}^{l}p_i(x)\prod_{j={l+1}}^{k}p_j(x)$ is $\sum_{j=0}^{r-t}\sum_{i=1}^{t}\binom{t}{i} \binom{r-t}{j}$.\\
\\To prove Step 10, we show that for each $1\leq i\leq \#S$ that,\\
ord$\left(p_i(x)^{a_i}\prod_{j\neq i}p_j(x)^{a_j}\right)=e$, where $p^{t_i-1}< a_i\leq p^t_i\ \left(t_i=\log_p{\frac{e}{\text{ord}(p_i(x))}}\right)$ and $0\leq a_j\leq f_j$.
\begin{align*}
    \text{ord}\left(p_i(x)^{a_i}\prod_{j\neq i}p_j(x)^{a_j}\right)
    &=\text{lcm}\left(\text{ord}\left(p_i(x)^{a_i}\right), \text{ord}\left(\prod_{j\neq i}p_j(x)^{a_j}\right)\right)\ \hspace{1 cm}(\text{by Theorem \ref{orderproduct}})\\
    &=e.\hspace{7.5cm}(\text{by Lemma \ref{for what powers exponent is e}})\\
\end{align*}
\begin{align*}
\text{The number of polynomials of the above form is}&\sum_{i=1}^{\#S}\sum_{j\neq i}(f_j+1)(p^{t_i}-p^{t_i-1})\\
    &=\sum_{i=1}^{\#S}\sum_{j\neq i}(f_j+1)\phi(p^{t_i}).\hspace{2cm} \square
\end{align*}
\begin{example}
There are exactly $4$ ternary cyclic codes of exponent $4$. In this case, the bound suggested by the algorithm is also $4$. In fact, $q=3, e=4$ and in $\mathbb{F}_3[x]$, $x^4-1=(x^2+1)(x+2)(x+1)$  so that 
$r=3, t=1$ and $\binom{2}{0}+\binom{2}{1}+\binom{2}{2}=4$.
\end{example}
\begin{example}
To find a lower bound for number of cyclic codes of exponent $12$ over $\mathbb{F}_3$. We factorize $x^{12}-1$ over $\mathbb{F}_3[x]$.\\
$x^{12}-1=(x+1)^3(x+2)^3(x^2+1)^3$.\\
Since gcd$(12,3)\neq 1$, we go to step 6 of the algorithm.\\
$N=\{p_1(x)=x+1, p_2(x)=x+2, p_3(x)=x^2+1\}$.\\\\
ord$(p_1(x))=2$, ord$(p_2(x))=1$, ord$(p_3(x))=4$\\\\
$k_1=\frac{12}{2}=6$, $k_2=\frac{12}{1}=12$, $k_3=\frac{12}{4}=3$\\\\
$S=\{p_3(x)\}$ and $t_3=\log_3{3}=1$.\\
Hence, the number of cyclic codes of exponent $12$ over $\mathbb{F}_3$ is atleast $(3+1)(3+1)\phi(3^1)=32$.
\end{example}
\section{Exponent of BCH codes}
\begin{definition}\cite{LING}\label{BCHcode}
Fix a primitive element $\beta$ in $\mathbb{F}_{q^m}$ and suppose $M^{(i)}(x)$ is the minimal polynomial of $\beta^i$ over $\mathbb{F}_q$. Let $a\in \mathbb{N}\cup\{0\}$ and let $\delta\leq q^m-1$. Denote by $g_{a,\delta}(x)$, the lcm of $M^{(i)}(x): a\leq i\leq a+\delta-2$. The cyclic code $C(a,\delta)$ generated by $g_{a,\delta}(x)$ over $\mathbb{F}_q$ is called the \emph{BCH code} of length $q^m-1$ with designed distance $\delta$. $g_{a,\delta}(x)$ will be simply denoted by $g(x)$ if $a$ and $\delta$ are clear from the context.
\end{definition}
\begin{lemma}\label{orderalpha}
Suppose $C_r$ denote the cyclotomic coset of $q$ in $\mathbb{Z}_{q^m-1}$ that contains $r$. If $C_r=C_s$, then ord$(\beta^s)=$ord$(\beta^r)$ for any primitive element $\beta$ of $\mathbb{F}_{q^m}$.
\end{lemma}
\textbf{Proof.} Suppose $s\in C_r$.
Then, $s=rq^j(\text{mod\ }q^m-1)$ for some $j\in \mathbb{N}\cup\{0\}$.\\
gcd$(s,q^m-1)=$gcd$(rq^j,q^m-1)=$gcd$(r,q^m-1)$ as gcd($q^j,q^m-1)=1.$\\
ord$(\beta^s)=\frac{q^m-1}{gcd(s,q^m-1)}=\frac{q^m-1}{gcd(r,q^m-1)}=$ord$(\beta^r)$. \hfill $\square$\\\\
We now give a formula for the exp$(C)$ in terms of the orders of $\beta^i$ in $\mathbb{F}_{q^m}^*$.
\begin{theorem}\label{ExponentBCH}
The exponent of the BCH code $C$ with its parameters as in Definition \ref{BCHcode} is given by
\begin{equation*}
    exp(C)=\text{lcm}\{\text{ord}(\beta^i): a\leq i\leq a+\delta-2\text\}, 
\end{equation*}
where ord$(\beta^i)$ denotes the order of $\beta^i$ in $\mathbb{F}_{q^m}^*$.
\end{theorem}
\textbf{Proof.} Let $n$ be the length of the BCH code. We have, $M^{(s)}(x)=M^{(t)}(x)$ iff $s$ and $t$ are in the same cyclotomic cosets of $q$ in $\mathbb{Z}_n$ by Corollary \ref{samecyclo}. Also $M^{(s)}(x)$ are irreducible over $\mathbb{F}_q$. Then the generator polynomial is
\begin{align*}
    g(x) = &\prod\{M^{(i)}(x):a\leq i\leq a+\delta-2\  \text{and}\ i \ \text{belongs to distinct cyclotomic cosets of}\\
    &  q\ \text{in}\ \mathbb{Z}_n\} 
\end{align*}
Therefore,
\begin{align*}
   \text{exp}(C)=&\text{ord}(g(x))\\
    =&\text{ord}(\prod\{M^{(i)}(x):a\leq i\leq a+\delta-2\  \text{and}\ i \ \text{belongs to distinct cyclotomic cosets of}\\
    q& \ \text{in}\  \mathbb{Z}_n\})\\
    =&\text{lcm}\{\text{ord}(M^{(i)}(x)):a\leq i\leq a+\delta-2\  \text{and}\ i \ \text{belongs to distinct cyclotomic cosets of}\\
    q& \ \text{in}\  \mathbb{Z}_n\}\hspace{5 cm} (\text{by Theorem \ref{orderproduct}})\\
    =&\text{lcm}\{\text{ord}(\beta^i):a\leq i\leq a+\delta-2\  \text{and}\ i \ \text{belongs to distinct cyclotomic cosets of}\\
    q& \ \text{in}\  \mathbb{Z}_n\} \hspace{5 cm}(\text{by Theorem \ref{orderirr}})\\
    =&\text{lcm}\{\text{ord}(\beta^i):a\leq i\leq a+\delta-2\  \text\} \hspace{.7cm}(\text{by Lemma \ref{orderalpha}})
\end{align*}
\normalsize
\hfill $\square$\\
\begin{corollary}
Let $\alpha, \beta \in \mathbb{F}_{q^m}$ be primitive elements. Denote by $M_1^{(i)}(x)$ and $M_2^{(i)}(x)$, the minimal polynomials of $\alpha^i$ and $\beta^i$ respectively over $\mathbb{F}_q$. Fix $a\in \mathbb{N}\cup\{0\}$ and designed distance $\delta$. Suppose $f(x)=lcm\{M_1^{(i)}(x): a\leq i\leq a+\delta-2\}$ and $g(x)=lcm\{M_2^{(i)}(x): a\leq i\leq a+\delta-2\}$. Let $C_\alpha$ and $C_\beta$ be the BCH codes of length $q^m-1$ generated by $f(x)$ and $g(x)$ respectively. Then $exp(C_\alpha)=exp(C_\beta)$.
\end{corollary}
\textbf{Proof.}
Since $\alpha$ and $\beta$ are two primitive elements of $\mathbb{F}_{q^m}$, therefore ord$(\alpha^i)$=ord$(\beta^i)$ $\forall i \in \mathbb{N}$. Then by the above theorem,
\begin{align*}
    \text{exp}(C_\alpha)&=\text{lcm}\{\text{ord}(\alpha^i): a\leq i\leq a+\delta-2\text\}\\
    &=\text{lcm}\{\text{ord}(\beta^i): a\leq i\leq a+\delta-2\text\}\\
    &=\text{exp}(C_\beta)
\end{align*}
\hfill $\square$
\begin{example}
Let $\beta$ be a root of $x^4+x+1=f(x)\in \mathbb{F}_2[x]$.  Since $f(x)$ is irreducible in $\mathbb{F}_2[x]$, $\beta$ is a primitive element of $\mathbb{F}_{16}\cong \frac{\mathbb{F}_2[x]}{<f(x)>}$.
\begin{enumerate}
\item[(i)] Denote by $C$, the $2$-ary $[15,9]$-BCH code generated by the lcm of $M^{(5)}(x)$ and $M^{(6)}(x)$ that is $(1+x+x^2)(1+x+x^2+x^3+x^4)$ whose designed distance is $3$. Then
\begin{align*}
   \text{exp}(C)=\text{lcm}(\text{ord}(\alpha^5),\text{ord}(\alpha^6))=\text{lcm}(3,5)=15.
\end{align*}
\item[(ii)] Denote by $D$, the $2$-ary $[15,13]$-BCH code generated by $M^{(5)}(x)=1+x+x^2$ and designed distance $2$. Then
\begin{align*}
   \text{exp}(D)=\text{lcm}(\text{ord}(\alpha^5))=3.
\end{align*}
\end{enumerate}
\end{example}
\begin{example}
Let $f(x)=2+x+x^2 \in \mathbb{F}_3[x]$ and let $\beta$ be its root. Since $f(x)$ is irreducible in $\mathbb{F}_3[x]$, $\beta$ is a primitive element of $\mathbb{F}_{9}\cong \frac{\mathbb{F}_3[x]}{<f(x)>}$.\\
The cyclotomic cosets of $3$ in $\mathbb{Z}_8$ are:\\
$C_0=\{0\}\hspace{3cm}C_1=\{1,3\}\hspace{3cm}C_2=\{2,6\}$\\
$C_4=\{4\}\hspace{3cm}C_5=\{5,7\}$
\begin{enumerate}
    \item[(i)] Denote by $C$, the $3$-ary narrow-sense $[8,4]$-BCH code which is generated by $g(x),$ the least common multiple of $M^{(1)}(x)$, $M^{(2)}(x)$ and $M^{(3)}(x)$ that is $(2+x+x^2)(1+x^2)$ whose designed distance is $4$. Then
\begin{align*}
   \text{exp}(C)=\text{lcm}(\text{ord}(\alpha^1), \text{ord}(\alpha^2), \text{ord}(\alpha^3))=\text{lcm}(8,4,8)=8.
\end{align*}
One may verify that $g(x)$ divides $x^8-1$ but does not divide $x^r-1$ for any $r<8$. Hence, exp($C$)=ord($g(x))=8$.
\item[(ii)]Denote by $D$, the $3$-ary $[8,6]$-BCH code which is generated by $g(x)=M^{(2)}(x)=(1+x^2)$ and designed distance $2$. Then
\begin{align*}
   \text{exp}(D)=\text{lcm}(\text{ord}(\alpha^2))=4.
\end{align*}
One may verify that $g(x)$ divides $x^4-1$ but does not divide $x^r-1$ for any $r<4$. Hence, exp($D$)=ord($g(x))=4$.
\end{enumerate}
\end{example}
From the above examples, one can observe that the exponent of BCH codes with designed distance greater than $2$ is equal to the length of the code. It is not an accident. In fact, we have:
\begin{corollary}
The exponent of a BCH code with designed distance greater than $2$ over $\mathbb{F}_q$ is equal to its length, namely, $q^m-1$.
\end{corollary}
\textbf{Proof.} Since $\delta>2$, $\{\text{ord}(\alpha^i): a\leq i\leq a+\delta-2\}$ contains at least two elements $\text{ord}(\alpha^a)$ and $\text{ord}(\alpha^{a+1})$.
Set $n=q^m-1, s=\gcd(n,a)$ and $r=\gcd(n,a+1)$ so that $\gcd(s,r)=1$. Then
\begin{align*}
    \text{exp}(C)&=\text{lcm}\{\text{ord}(\alpha^i): a\leq i\leq a+\delta-2\text\}\\
    &\geq\text{lcm}(\text{ord}(\alpha^a),\text{ord}(\alpha^{a+1}))\\
    &=\text{lcm}(\frac{n}{s}, \frac{n}{r})\\
    &=\frac{n}{\gcd(s,r)}\\
    &=n.
\end{align*}
Thus, exp$(C)\geq n$. But by definition exp$(C)\leq n$, proving that exp$(C)=n=q^m-1$.\hfill $\square$
\subsection{Exponent of Reed-Solomon code}
\begin{definition}\cite{LING}\label{RScode}
The BCH code generated by $(x-\beta^{a+1})(x-\beta^{a+2})...(x-\beta^{a+\delta-1})\in \mathbb{F}_q[x]$, where $a\in \mathbb{N}\cup\{0\}, 2\leq \delta \leq q-1$, and $\beta \in \mathbb{F}_{q}$ is primitive, is called a $q$-ary \emph{Reed-Solomon code} (RS code).
\end{definition}
As RS codes are BCH codes, Theorem \ref{ExponentBCH} can be rephrased as follows:
\begin{theorem}
The exponent of the RS code $C$ with its parameters as in Definition \ref{RScode} is given by
\begin{equation*}
    \text{exp}(C)=\text{lcm}\{\text{ord}(\beta^i): a+1\leq i\leq a+\delta-1\text\}, 
\end{equation*}
where ord$(\beta^i)$ denotes the order of $\beta^i$ in $\mathbb{F}_{q}^*$.

\end{theorem}
\begin{corollary}
The exponent of a RS code with designed distance greater than $2$ over $\mathbb{F}_q$ is equal to its length, namely, $q-1$.

\end{corollary}
\section{Conclusion}
In this article, we find bounds for the number of cyclic codes of given exponent over every finite field. We also determine a formula for the exponent of the BCH code and RS code and conclude that their exponent is same as the length of the code, if the designed distance is greater than $2$. For a cyclic code, the exponent is believed to have certain significance to the code, so in future one can study the importance of the exponent of a cyclic code.

\bibliographystyle{abbrv}
\bibliography{exponent}

\begin{thebibliography}{1}

\bibitem{ANNA}
N.~Annamalai and C.~Durairajan.
\newblock Exponent of cyclic codes over.
\newblock {\em Journal of Discrete Mathematical Sciences and Cryptography},
  0(0):1--6, 2021.

\bibitem{FIELD}
R.~Lidl and H.~Niederreiter.
\newblock {\em Introduction to finite fields and their applications}.
\newblock Cambridge University Press, Cambridge, first edition, 1994.

\bibitem{LING}
S.~Ling and C.~Xing.
\newblock {\em Coding theory}.
\newblock Cambridge University Press, Cambridge, 2004.
\newblock A first course.

\end{thebibliography}
\end{document}